\documentclass{article}
\usepackage[dvips]{graphicx}

\title{Quark masses, the Dashen phase, and gauge field topology}

\author{Michael Creutz \\
Brookhaven National Laboratory, \\
Upton, NY 11973, USA
\thanks{ This manuscript has been authored under contract
    number DE-AC02-98CH10886 with the U.S.~Department of Energy.
    Accordingly, the U.S. Government retains a non-exclusive,
    royalty-free license to publish or reproduce the published form of
    this contribution, or allow others to do so, for U.S.~Government
    purposes.  }
}
	
\date{July 2013}

\begin{document}

\maketitle

\begin{abstract}
The CP violating Dashen phase in QCD is predicted by chiral
perturbation theory to occur when the up-down quark mass difference
becomes sufficiently large at fixed down-quark mass.  Before reaching
this phase, all physical hadronic masses and scattering amplitudes are
expected to behave smoothly with the up-quark mass, even as this mass
passes through zero.  In Euclidean space, the topological
susceptibility of the gauge fields is positive at positive quark
masses but diverges to negative infinity as the Dashen phase is
approached.  A zero in this susceptibility provides a tentative signal
for the point where the mass of the up quark vanishes.  I discuss
potential ambiguities with this determination.

\end{abstract}

\section{Introduction}

Because of confinement, properties of quarks such as their masses
cannot be observed directly.  Indeed, what does a quark mass mean when
free quarks do n0t exist?  The commonly advocated approach is to do a
lattice calculation and adjust the bare quark masses to match physical
hadron properties.  As one takes the continuum limit, the bare quark
masses will flow as determined by the renormalization group.  Because
of asymptotic freedom and given a renormalization scheme, this
behavior is precisely determined and allows us to extract a
renormalized quark mass.  This procedure is briefly reviewed in
Appendix A.  However, the details can depend on the precise lattice
formulation, which raises the possibility of an ambiguity in defining
the massless quark limit.

In a letter a number of years ago I pointed out that non-perturbative
effects can leave an additive ambiguity in the definition of a
non-degenerate quark mass \cite{Creutz:2003xc}.  This additive effect
goes away in the isospin limit, a limit in which most lattice
calculations are currently done.  In the light quark regime, the quark
masses are closely tied to the pseudo-scalar meson spectrum, with
massless pions appearing as the up and down quarks become massless
together.

Of course isospin in nature is only an approximate symmetry, being
broken by the non-degeneracy of the up and down quarks as well as by
electromagnetism.  Indeed, it is somewhat remarkable that both effects
are of comparable order to the hadron spectrum.  The charged and
neutral pion mass difference of a few MeV is generally regarded as
being primarily from electromagnetic effects while the neutron-proton
mass difference receives comparable contributions both from an
underlying quark mass difference and from electromagnetism.

In chiral perturbation theory, to lowest order the pions have a mass
squared proportional to the average of the up and down-quark masses.
With isospin breaking, the neutral and charged pions are no longer
exactly degenerate.  The dominant effect is the energy in the
electromagnetic field of the charged pion, leaving the neutral pion
the lightest of the three.  In addition, chiral perturbation theory
predicts a small further splitting proportional to the square of the
up-down mass difference \cite{Creutz:1995wf}.  Here I will concentrate
on the quark mass difference effects, although presumably the
electromagnetic effects have similar consequences and do not modify
the qualitative picture below.

Since the pion mass is tied to the average of the up and down quark
masses, there remains a gap in the hadronic spectrum if the down quark
remains massive while the up-quark mass is taken to zero.  Indeed, all
physical processes are expected to behave smoothly in this limit.
This brings us back to the question in Ref.~\cite{Creutz:2003xc} of
whether there is a precise meaning to having a vanishing up-quark mass.

In this paper I discuss the qualitative behavior of the theory in the
vicinity of vanishing up-quark mass, with particular attention to
topological issues with the gauge fields.  Topology is important since
the pion mass difference receives a non-perturbative contribution
through the induced mixing of the pion and the eta prime mesons, and
the eta prime mass is dominated by non-perturbative effects from
topology \cite{'tHooft:1976fv}.  This mixing behaves smoothly even at
a zero up-quark mass, despite the classical suppression of topological
effects from a zero in the fermion determinant at non-trivial
topology.  In particular the eta prime mass, which comes primarily
from these effects, remains of order the strong interaction scale
throughout this region.

Continuing to negative up-quark mass, I argue that the topological
susceptibility will diverge to negative infinity as one approaches
what is known as the Dashen phase \cite{Dashen:1970et}.  Since this
susceptibility is a positive quantity for positive quark masses, it
must show a zero before this divergence becomes dominant.  Assuming a
single zero, this provides a natural definition of the point of
vanishing up-quark mass.  Nevertheless, since typical configurations
in a path integral are non-differentiable, there may be subtleties
that can lead to ambiguities in defining topology.

In Section \ref{spectrum} I review the standard picture from chiral
perturbation theory of how the three pions are degenerate up to second
order in the quark masses.  In Section \ref{dashen} I discuss the
continuation to negative up-quark masses and why the Dashen phase is
expected to appear.  Section \ref{topology} discusses the topological
susceptibility as a signal of where the up-quark mass vanishes.  Here
I argue for the divergence of this quantity as the Dashen phase is
approached.  Then in Section \ref{lattice} I discuss the ambiguities
that can arise in trying to define topology on the lattice.  Section
\ref{symmetries} discusses how these ambiguities are not resolved by
through current algebra and the symmetries of the theory.  A brief
summary of the conclusions appears in Section \ref{conclusions}.

A couple of closely related topics are relegated to appendices.
Appendix A reviews how the bare quark mass always flows to zero in the
continuum limit and how to define a renormalized mass from this flow.
Appendix B discusses how the absence of a CP violating Theta parameter
dependence of QCD at vanishing mass is a tautology associated with
using singular coordinates.

\section{The pseudo-scalar spectrum and the anomaly}
\label{spectrum}

In this section I review the standard picture of the pseudo-scalar
meson spectrum with two flavors of light fermions, the $u$ and $d$
quarks.  From the fermion fields I can construct four pseudo-scalar
operators
\begin{eqnarray}
\overline u \gamma_5 d &\sim& \pi_+ \cr
\overline d \gamma_5 u &\sim& \pi_- \cr
\overline u \gamma_5 u &&  \cr
\overline d \gamma_5 d &&  
\end{eqnarray}
The first two create the charged pions, which have a natural mass
controlled by the average of the up and down quark masses.

\begin{figure}
\centering
\includegraphics[width=3in]{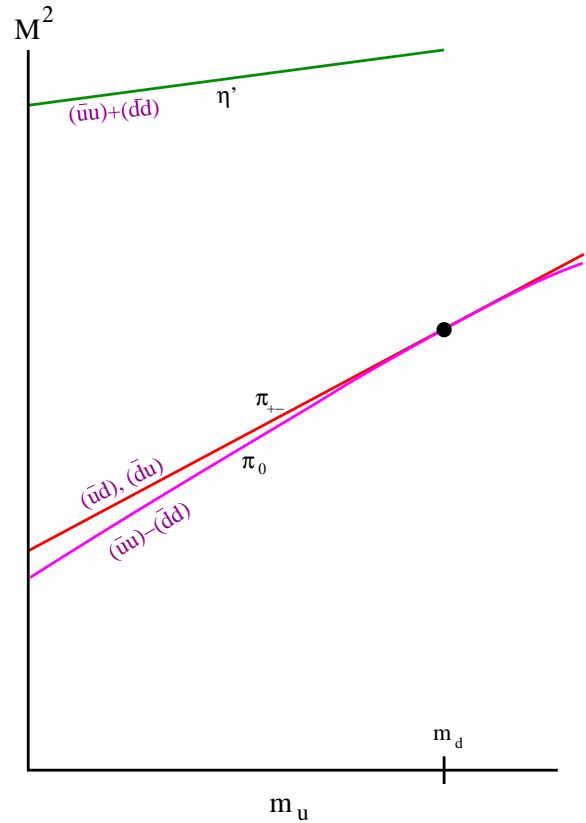}
\caption{The qualitative behavior of the meson spectrum as a function
  of the up-quark mass with a fixed non-vanishing down-quark mass.
  Note that a mass gap persists as the up-quark mass approaches zero.
  Away from the isospin limit, mixing of the neutral pion with the eta
  prime can result in splitting between the neutral and charged pions.
  This splitting is quadratic in the quark mass difference.  }
\label{fig1}
\end{figure}

All of these operators involve a helicity flip.  Helicity conservation
is usually taken as a property of gauge theories coupled to light
fermions.  Thus for light quarks one might naively argue that the last
two operators above would not mix significantly as that would involve
helicity change for both the up and down quarks.  This suggests that
there would be two light neutral pseudo-scalars, one whose mass is
controlled by the up-quark mass and a second by the down-quark mass.

Of course this expectation is naive.  The anomaly induces a strong
mixing between the neutral pseudo-scalar states.  The symmetric
combination
\begin{equation}
\overline u \gamma_5 u+\overline d \gamma_5 d\sim \eta^\prime
\end{equation}
 gains a mass of order $\Lambda_{qcd}$ and becomes the eta prime
 meson.  How this mixing takes place through gauge configurations of
 non-trivial topology was elegantly described by 't Hooft
 \cite{'tHooft:1976fv}, and the mixing term is often referred to as
 the 't Hooft vertex.

After the above mixing takes place, one is left with the orthogonal
combination
\begin{equation}
\overline u \gamma_5 u-\overline d \gamma_5 d\sim\pi_0, 
\end{equation}
to represents the neutral pion.  Having approximately equal
contributions from each quark, the $\pi_0$ also has its mass
dominantly controlled by the average quark mass.  In the isospin limit
of equal up and down quark masses, the three pions are degenerate up
to electromagnetic splittings, which I ignore here.

Away from the isospin limit, small mixings of the eta prime and the
neutral pion should remain.  This allows for a purely hadronic
contribution to the pion mass splitting proportional to $(m_d-m_u)^2$.
The sign of this term is generally expected to drive the neutral pion
mass down, although symmetry alone does not determine this.  The
qualitative picture is sketched in Fig.~(\ref{fig1}).

\section{The Dashen phase}
\label{dashen}

An important observation is that when the up-quark mass vanishes with
a non-vanishing down-quark mass, the theory retains a mass gap.  This
leads to the question of what happens as the up-quark mass is varied
to negative values.  The quark mass itself is only a formal parameter
while the lightest physical particles are the pions.  Their masses are
controlled primarily by the average of the up and down quark masses,
and it is natural to expect them to continue to drop until the
up-quark mass is comparable in magnitude to the negative of the
down-quark mass.  The behavior of all physical quantities should be
smooth in the quark masses in the immediate vicinity of vanishing
up-quark mass as long as the down-quark mass remains non-zero.

\begin{figure}
\centering
\includegraphics[width=4in]{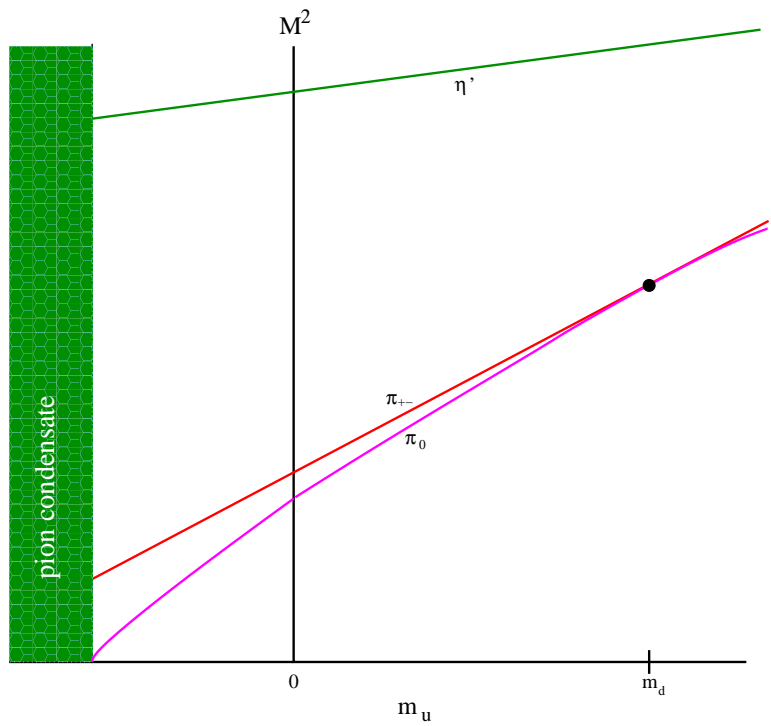}
\caption{Continuing the up-quark mass into the negative regime brings
  up the possibility of a vanishing pion mass and a subsequent
  condensation of the neutral pion field. 
}
\label{fig2}
\end{figure}

On continuing to decrease the up-quark mass further into the negative
region, the neutral pion mass will become progressively lighter and
may eventually hit zero.  Beyond such a point, the neutral pions
should form a condensate with the pion field acquiring an expectation
value.  As the pion is CP odd, the new phase spontaneously breaks CP
symmetry.  The possibility of such a spontaneous breaking was
postulated some time ago by Dashen \cite{Dashen:1970et}.  As that was
before the development of QCD as the theory underlying the strong
interactions, his argument was based on current algebra ideas.  The
qualitative picture is illustrated in Fig.~(\ref{fig2}).

Note that this phenomenon is inherently non-perturbative.  Naively the
sign of a fermion mass is irrelevant in perturbative diagrams because
it can be reversed by a chiral rotation.  That rotation, however, is
not valid due the chiral anomaly.  This pion condensation occurs in a
region where the product of the quark masses is negative.  As is well
known and discussed briefly in Appendix B, when the quark masses are
made complex, the phase $\Theta$ of the product of the quark masses is
a physical parameter \cite{'tHooft:1976fv}.  Thus the Dashen phase
occurs in a region where formally $\Theta=\pi$.  But note from this
example that there is also a finite region of negative up-quark mass
before this condensation takes place.  Thus whether or not there is a
spontaneous breaking of CP at $\Theta=\pi$ will depend on the detailed
values of the quark masses.

Vafa and Witten \cite{Vafa:1984xg} have argued that QCD cannot
spontaneously break CP.  This argument assumed that one is working in
a region where the fermion determinant appearing in the path integral
is strictly positive, {\sl i.e.} at $\Theta=0$.  The Dashen phase is
not a counterexample to their argument since it occurs where on some
gauge field configurations the up-quark determinant is negative.

\section{Implications for the path integral }
\label{topology}

The above discussion of how the physical states behave as the quark
masses are varied implicitly applies to the properties of the
Minkowski space theory.  The picture raises some interesting questions
on how this physics appears in the path integral formulation.  In
particular, when fermions are integrated out in terms of the
determinant of the Dirac operator, what are the implications for the
behavior of this determinant?

Simple chiral Lagrangian arguments suggest that the boundary of the
Dashen phase involves a second order phase transition \cite{
  Creutz:2003xu, Creutz:2010ts}.  At the critical point, the divergent
correlation length is associated with the vanishing of the neutral
pion mass.  This example shows explicitly that one can have vanishing
particle masses or divergent correlations in QCD at a point where none
of the quark masses vanish.  Because of the mass term, the Dirac
operator will not have any small or vanishing eigenvalues.  

In contrast, the fact that a mass gap persists in the vicinity of
vanishing up-quark mass is an explicit example of a situation where
the Dirac operator could have small eigenvalues but this does not
imply important long range physics.  Furthermore, the absence of any
obvious structure raises the question of whether there is some
physical way to determine exactly where the up-quark mass vanishes.
That was the question raised in \cite{Creutz:2003xc}.

In the path integral approach, one standard answer to defining a
vanishing up-quark mass is that this represents the point where the
topological susceptibility vanishes.  This is connected to the index
theorem, which shows that the massless Dirac operator has a zero
eigenvalue whenever the gauge field has non-trivial topology.  

In any given gauge configuration the topological charge is formally
defined as
\begin{equation}
\nu={g^2\over 16 \pi^2}\int d^4x F_{\mu\nu}(x)\tilde F_{\mu\nu}(x).
\end{equation}
The susceptibility then follows from the path integral as
\begin{equation}
\xi= {1\over V}\langle \nu^2 \rangle
\end{equation}
where $V$ is the space-time volume and the expectation is over the
space of gauge configurations appropriately weighted with the action,
including the fermion determinant.  Using translational invariance, 
this can be written
\begin{equation}
\label{xi}
\xi= 
\left({g^2\over 16 \pi^2}\right)^2 \left \langle\int d^4x 
F_{\mu\nu}(x)\tilde F_{\mu\nu}(x)F_{\mu\nu}(0)\tilde F_{\mu\nu}(0)
\right \rangle.
\end{equation}
For convenience below I abbreviate $F_{\mu\nu}(x)\tilde F_{\mu\nu}(x)$
as $F\tilde F(x)$.

It is important to note that since $F\tilde F$ is an odd operator under
time reversal, the correlator
\begin{equation}
\label{correlator}
\langle F\tilde F(x) F\tilde F(0) \rangle
\end{equation}
is negative at non-zero separations \cite{Seiler:2001je}.  This is
true even at positive quark masses where the path integral weight is
strictly positive.  Since the full susceptibility $\xi$ is defined as
a square, it is expected to be positive when the quark masses are
positive.  The interpretation is that the correlator in
Eq. (\ref{correlator}) has a singular contact term providing a
positive contribution that exceeds the negative part from non-zero
separation.  We will shortly see that the situation is more
complicated when the up-quark mass is negative.

It is straightforward to show that as a function of the up-quark mass
the topological susceptibility must vanish somewhere before reaching
the Dashen phase.  At negative quark mass the susceptibility, despite
being formally a square, need not be positive.  This is possible
because the fermion determinant that enters the weight for the path
integral can be negative.  Although the path integral then loses its
probability interpretation, presumably it still exists as a correct
approach to the quantum theory.  

For a negative up-quark mass near zero one expects higher topology
configurations to be strongly suppressed.  In this situation, the
topological susceptibility should be dominated by the minimal
non-trivial winding number of unity (winding number zero gives no
contribution to the susceptibility), and in this case the fermion
determinant is negative.

Not only can $\xi$ be negative, but as one approaches the endpoint of
the Dashen phase, the topological susceptibility actually diverges to
negative infinity.  Since for positive mass the susceptibility is a
positive definite quantity, there must be a zero somewhere before we
reach the Dashen phase.  The position of this zero is a natural
definition of theory with zero up-quark mass.

The divergence of the susceptibility at the boundary of the Dashen
phase is a direct consequence of the vanishing of the neutral pion
mass at that point.  Since isospin is broken, one expects some mixing
between the neutral pion and both the eta prime meson and any
pseudo-scalar glueball states.  In particular, the operator $F\tilde F$
should have a finite amplitude to create the neutral pion; {\it i.e.}
$\langle \pi_0 | F\tilde F | 0\rangle \ne 0$.  This means that at long
distances the integral in Eq.~(\ref{xi}) will diverge as
$-1/M_{\pi_0}^2$ due to pion intermediate states .  Again, since
$F\tilde F$ is odd under time reversal, this divergence is to negative
infinity \cite{Seiler:2001je}.  And since the boundary of the Dashen
phase is determined by long distance pion dynamics, one does not
expect the contact term to have a corresponding divergence, nor does
one expect the coupling of $F\tilde F$ to the pion to vanish.

\section{Topology and the lattice}
\label{lattice}

So we see that as the up-quark mass varies, there must be a point
where the topological susceptibility vanishes.  But is this actually a
physical concept, or could it depend on how the theory is regulated?
This is a non-trivial point since in a path integral typical gauge
configurations involve non-differentiable fields.  Since gauge field
topology is a non-perturbative concept, one must consider this
question in the context of a non-perturbative regulator, {\it i.e.}
the lattice.  Indeed, numerous lattice schemes for defining topology
have been proposed, but all presented so far leave room for
ambiguities.  This is actually a rather old topic; some numerical
studies appear in
Refs. \cite{Teper:1985rb,Bruckmann:2009cv,Moran:2010rn}.

One could just take a simple discretization of $F\tilde F$ and sum its
value over configurations taken from a Monte Carlo simulation.  One
example of this is in Ref. \cite{Creutz:2010ec}.  The problem with
this approach is that unlike with smooth continuum fields, any local
definition of $F\tilde F$ will not in general be a total derivative
and its integral will not be an integer.  As such it is not a true
topological object.

To get around this one might impose a smoothness condition on the
lattice fields such that one can uniquely construct continuum field
interpolating between the lattice sites \cite{Luscher:1981zq,
  Adams:1998eg}.  This leaves open a few questions.  First, a
constraint on the lattice action that is not analytic in the gauge
field variables will lead to a non-positive transfer matrix and
correspondingly a non-Hermitean Hamiltonian when the cutoff is in
place \cite{Creutz:2004ir}.  Second, as one takes the continuum limit,
it is unclear to what extent different smoothness conditions will give
the same result.

As a variation on the previous, one can perform a smoothing on the
gauge field until it settles into a state of well defined topology.
One example is to use a differential flow with the Wilson gauge action
as a potential \cite{Narayanan:2006rf,Luscher:2010iy}.  This
procedure, however, is non-unique.  First, the result can depend on
the cooling action used.  The standard Wilson gauge action is one
choice, but it is unclear why this should be the appropriate form when
effects of dynamical quarks are present.  Second, the winding number
found can depend on how long one cools.  Without some constraint on
the local action, topological objects can shrink with cooling and
``fall through the lattice.''

This leads us to a class of prescriptions based on the index theorem
relating zero modes of the Dirac operator to the gauge field topology
\cite{Adams:2000rn}.  The simplest is to count the small real
eigenvalues of the Wilson Dirac operator.  The ambiguity here lies in
the definition of ``small.''  If one takes all real eigenvalues and
counts them with their chirality, one will always get zero.
Presumably one should ignore the eigenvalues in the doubler region,
but with dynamical fermions the border between physical and doubler
modes becomes blurred.  Note that the zero of the topological
susceptibility must occur for the up-quark mass in a region where
negative real eigenvalues can occur.  This allows a cancellation of
the purely positive contribution when there are no negative real
eigenvalues against the negative contribution when such modes do
exist.

An action satisfying the Ginsparg-Wilson relation
\cite{Ginsparg:1981bj} can have exact zero modes and they properly
match with the gauge field index when the gauge fields are smooth.
With such an approach, a vanishing up quark mass automatically
suppresses all configurations of non-zero topology since the fermion
determinant is zero for any such.  This, however, does not resolve the
issue because there are many operators that satisfy this relation, and
for general fields they do not all give the same index.  The Neuberger
construction of the overlap operator \cite{Neuberger:1997fp} depends
on a projection from a kernel such as the Wilson operator.  The
location of the projection point suffers the same ambiguity as in the
previous paragraph.  Indeed, one would expect the eigenvalue count for
typical configurations to decrease as the projection point is lowered
and fewer real eigenvalues of the Wilson kernel lie below this
location.  Again, the question here is tied to the real eigenvalues,
which are inherently non-perturbative.  It is often asserted that with
the Wilson kernel there will be a region where the density of real
eigenvalues will decrease rapidly enough to eliminate this ambiguity,
although the numerical evidence for this remains limited
\cite{Edwards:1998wx,Giusti:2004qd,Luscher:2010ik}.

I note in passing that the staggered fermion approach to lattice
fermions retains the naive symmetry under reversing the sign of any
quark mass.  This ensures the uniqueness of the zero mass theory.
However this approach suffers doubling issues, making any massless
species actually multiply degenerate.  The rooting procedure often
advocated to circumvent this problem is known to be an approximation
which severely mutilates the 't Hooft vertex
\cite{Creutz:2007yg,Creutz:2011hy}.  Unlike as discussed in Section
\ref{spectrum}, taste non-singlet $\overline u \gamma_5 u$ mesons will
survive the procedure with masses controlled by the up-quark mass
alone.  Pairs of these spurious mesons will cause unphysical
thresholds in scattering amplitudes when the physical quarks are not
degenerate, even when all quarks have positive mass.

\section{Symmetries and currents}
\label{symmetries}

One might try to circumvent these issues via continuum concepts using
currents related to the symmetries of the theory.  Indeed, Dashen's
discussion of the possibility of a CP violating phase was based on
current algebra and was presented before QCD became the accepted
theory for the strong interactions.  In this section I will discuss
how the lack of any symmetry at vanishing up quark mass leaves open an
ambiguity in defining this point.

When the quark mass difference vanishes we have isospin symmetry.
This makes a zero mass difference a well defined concept.  To get a
handle on the the quark mass difference as we move away from this
point, one might consider the partial conservation of a charged vector
current.  For a small mass difference the divergence of such a current
is proportional to the mass difference with a coefficient depending on
the chiral condensate.  However, as the mass difference grows, the
behavior of this divergence ceases to be linear.  In particular, as
one passes through the boundary of the Dashen phase, an associated
non-analyticity will appear in all physical quantities, including the
charged currents.  It is these non-linearities that leave room for the
ambiguities discussed in the previous section.

Another interesting situation occurs when the sum of the up and down
quark masses vanishes.  Then, when the quark mass difference does not
vanish, we are deep in the Dashen phase.  Here we can move the sign of
the negative quark mass into the gauge field topology with a term in
the action proportional to $F\tilde F$.  Then there is again an
explicit isospin symmetry between the quarks.  This serves to protect
the mass sum from renormalization.  We again have a symmetry that
protects the average quark mass from an additive renormalization.

The issue of a vanishing up quark mass when the down quark is massive
is not associated with either of these two symmetry points.  Rather,
it involves neither the sum nor the difference of the quark masses
vanishing.  In this case these quantities are not related by any
fundamental symmetry.  For example, they transform differently under
isospin; the sum of the masses is an isoscalar quantity and the
difference is an isovector.  While both are individually
multiplicatively renormalized, there is no rigorous symmetry that
relates their non-perturbative renormalization factors.

A further complication is that non-perturbative physics requires a
lattice approach, and then properly defined currents are dependent on
the detailed fermion formulation.  With all lattice actions, the
conserved currents are not simple on-site quantities.  Particularly
with Ginsparg-Wilson type operators, they involve fields spread over
formally unlimited distances.  Given a particular lattice formulation,
the underlying quark masses may be well defined, but there appears to
be no fundamental reason to require the physics at a vanishing quark
mass to be universal between schemes.

\section{Conclusions}
\label{conclusions}

There are three main conclusions to draw from this discussion.  First,
the topological susceptibility must diverge as the Dashen phase is
approached.  Second, since this divergence is to negative infinity,
with any sensible regulator there must exist a zero in the
susceptibility as the up quark mass varies up to its physical value.
Third, it is unclear whether the location of this zero is universal
between lattice schemes.

The third point is somewhat controversial since this zero provides a
tentative definition of the point where the up quark is massless.
Because all known methods to locate this point in a lattice simulation
appear to have some arbitrariness, it remains unclear whether the zero
must scale in a manner to give a unique continuum limit for more
physical quantities.  In particular, two different lattice cutoffs
taken to the continuum limit while forcing $m_u=0$ can potentially
give different ratios for physical hadron masses.

From a phenomenological point of view, there seems to be little reason
to care whether the up-quark mass is ambiguous.  This is not a
directly observable quantity, and both hadron masses and scattering
amplitudes behave smoothly as the quark mass passes through zero.  For
confidence in lattice results, it is important to compare physical
results with different schemes.  The issues raised here suggest that
using quark masses or topological susceptibility directly for such
matching might lead to unpredictable results.

\section*{Appendix A: Quark masses and the renormalization group}

In lattice gauge language, asymptotic freedom tells us how to vary the bare
gauge coupling and quark masses to take the continuum limit
This variation is manifest via the ``renormalization group equations'' 
\begin{eqnarray}
&&a{dg\over da}=\beta(g)=\beta_0 g^3+\beta_1 g^5 +\ldots
+{\rm non{\hbox{-}}perturbative}\cr
&&a{dm\over da}=m\gamma(g)=m(\gamma_0 g^2+\gamma_1 g^4 +\ldots)
+{\rm non{\hbox{-}}perturbative.}
\label{rgroup}
\end{eqnarray}
Here the three initial perturbative coefficients
$\beta_0,\ \beta_1,\ \gamma_0$ are scheme independent and known
\cite{Politzer:1973fx,Gross:1973id,Gross:1973ju,Caswell:1974gg,Jones:1974mm,Vermaseren:1997fq,Chetyrkin:1997fm}.
These equations are easily integrated to show
\begin{eqnarray}
&&a={1\over \Lambda} e^{-1/2\beta_0 g^2} g^{-\beta_1/\beta_0^2}
(1+O(g^2))\cr
&&m=Mg^{\gamma_0/\beta_0}
(1+O(g^2)).
\end{eqnarray}
The quantities $\Lambda$ and $M$ are ``integration constants'' for the
renormalization group equations.  Rewriting these relations gives the
coupling and mass flow in the continuum limit $a\rightarrow 0$
\begin{eqnarray}
&&g^2\sim {1\over \log(1/\Lambda a)}\rightarrow 0\qquad\qquad
\hbox{``asymptotic freedom''}\cr
&&m\sim M\ \left({1\over \log(1/\Lambda
a)}\right)^{\gamma_0/\beta_0}\rightarrow 0.
\end{eqnarray}
Here $\Lambda$ is usually regarded as the ``QCD scale'' and $M$ as the
``renormalized quark mass.''  

Thus it would seem that given some renormalization scheme we have a
way to define the quark mass.  The issue with defining a vanishing
quark mass is complicated due to the ``non-perturbative'' terms in
Eq.~(\ref{rgroup}).  Because the integration constant $\Lambda$
introduces a new scale into the theory, the non-perturbative
contribution to the mass flow need not be proportional to the bare
mass.  Without some symmetry to prevent it, an additive mass shift
proportional to $\Lambda$ is generally possible.  And with only one
massless quark, the anomaly breaks the naive chiral symmetry that
would prevent such a term.

Of course with multiple degenerate quarks we would have flavor
non-singlet chiral symmetries that would remove any such additive
renormalization.  But in the case considered here, we are keeping the
down-quark mass fixed at a non-zero value.  The fact that one would
recover chiral symmetry if the down-quark mass was also zero
indicates that the size of any potential ambiguity in the up-quark
mass should be proportional to the down-quark mass.

\section*{Appendix B: The Theta parameter and CP symmetry}

As is well known, QCD as a renormalizable quantum field theory has the
possibility of including a CP violating term, usually called $\Theta$.
This is closely tied to the anomaly, and can be moved around between
appearing either in the quark mass matrix or the appearance of a
topological term in the gauge field action.  One often proposed way to
locate the vanishing up-quark mass point is that at this point the
theory is independent of the Theta parameter.  But in the light of the
above discussion this is actually a tautology.  

After rotating the theta parameter into the up-quark mass, this mass
term takes the form
\begin{equation}
m_1 \overline \psi \psi +im_5 \overline \psi\gamma_5\psi.
\end{equation}
From this, it is conventional to define a complex mass parameter
\begin{equation}
m=m_1+im_5.
\end{equation}
The phase of this quantity is the parameter $\Theta$.  Of course, if
$m$ vanishes as a complex number, $\Theta$ ill defined and irrelevant.

The problem with this interpretation is that $m_1$ and $m_5$ are
independent parameters.  And the above discussion suggested an
ambiguity in defining where $m_1$ vanishes.  This ambiguity feeds
through into an ambiguity in defining $\Theta$.  In a sense the
conventional approach attempts to set up polar coordinates about a
point which is not a natural origin.


\end{document}